%% file: main.tex
\documentclass[sigconf]{acmart}
\pdfoutput=1

\usepackage{xcolor} 
\usepackage{balance}
\usepackage{hyperref}

\copyrightyear{2024}
\acmYear{2024}
\setcopyright{rightsretained}
\acmConference[SIGCSE 2024]{Proceedings of the 55th ACM Technical Symposium on Computer Science Education V. 1}{March 20--23, 2024}{Portland, OR, USA}
\acmBooktitle{Proceedings of the 55th ACM Technical Symposium on Computer Science Education V. 1 (SIGCSE 2024), March 20--23, 2024, Portland, OR, USA}
\acmDOI{10.1145/3626252.3630795}
\acmISBN{979-8-4007-0423-9/24/03}

\makeatletter
\gdef\@copyrightpermission{
  \begin{minipage}{0.3\columnwidth}
    \href{https://creativecommons.org/licenses/by/4.0/}{\includegraphics[width=0.90\textwidth]{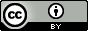}}
  \end{minipage}\hfill
  \begin{minipage}{0.7\columnwidth}
    \href{https://creativecommons.org/licenses/by/4.0/}{This work is licensed under a Creative Commons Attribution International 4.0 License.}
  \end{minipage}
  \vspace{5pt}
}
\makeatother

\begin{document}

\settopmatter{printacmref=true}

\title{Bringing Social Computing to Secondary School Classrooms}

\author{Kianna Bolante}
\email{kbolan@cs.washington.edu}
\affiliation{%
  \institution{University of Washington}
  \city{Seattle}
  \state{Washington}
  \country{USA}
}

\author{Kevin Chen}
\email{kc@cs.washington.edu}
\affiliation{%
  \institution{Cleveland STEM High School}
  \city{Seattle}
  \state{Washington}
  \country{USA}}

\author{Quan Ze Chen}
\email{cqz@cs.washington.edu}
\affiliation{%
  \institution{University of Washington}
  \city{Seattle}
  \state{Washington}
  \country{USA}
}

\author{Amy Zhang}
\email{axz@cs.uw.edu}
\affiliation{%
  \institution{University of Washington}
  \city{Seattle}
  \state{Washington}
  \country{USA}
}

\renewcommand{\shortauthors}{Kianna Bolante, Kevin Chen, Quan Ze Chen, \& Amy Zhang}


\input{my-files/abstract}



\maketitle
\section{Introduction}
\input{my-files/introduction}

\section{Related Work}
\input{my-files/related-work}

\section{Lesson Development}
\input{my-files/lessons}

\section{School Visits}
\input{my-files/visits}

\section{Results}
\input{my-files/results}

\section{Teaching Resources}
\input{my-files/teaching-resources}

\section{Reflection}
\input{my-files/reflection}

\section{Next Steps}
\input{my-files/next-steps}

\begin{acks}
We extend our gratitude to the staff and faculty at Seattle Public Schools who contributed to outreach initiatives and supported school visits in the interest of social computing education. 
\end{acks}

\bibliographystyle{ACM-Reference-Format}
\balance
\bibliography{sample-base}

\end{document}

%% file: my-files/abstract.tex
\begin{abstract}
Social computing is the study of how technology shapes human social interactions. This topic has become increasingly relevant to secondary school students (ages 11--18) as more of young people's everyday social experiences take place online, particularly with the continuing effects of the COVID-19 pandemic. However, social computing topics are rarely touched upon in existing middle and high school curricula. We seek to introduce concepts from social computing to secondary school students so they can understand how computing has wide-ranging social implications that touch upon their everyday lives, as well as think critically about both the positive and negative sides of different social technology designs.

In this report, we present a series of six lessons combining presentations and hands-on activities covering topics within social computing and detail our experience teaching these lessons to approximately 1,405 students across 13 middle and high schools in our local school district. We developed lessons covering how social computing relates to the topics of Data Management, Encrypted Messaging, Human-Computer Interaction Careers, Machine Learning and Bias, Misinformation, and Online Behavior. We found that 81.13\% of students expressed greater interest in the content of our lessons compared to their interest in STEM overall. We also found from pre- and post-lesson comprehension questions that 63.65\% learned new concepts from the main activity. We release all lesson materials on a website for public use. From our experience, we observed that students were engaged in these topics and found enjoyment in finding connections between computing and their own lives.
\end{abstract}

%% file: my-files/introduction.tex
In today's rapidly evolving digital world along with the impact that COVID-19 has had on technology use, social computing platforms have become an integral part of the lives of young individuals \cite{branscombe2020network}. It profoundly influences how they communicate, learn, and interact with the world around them. The pervasiveness of social media, online communities, and virtual spaces has not only transformed the way information is disseminated but has also presented unique challenges and opportunities for society at large. As the younger generation navigates this dynamic technological space, there is a growing imperative to equip them with the necessary skills, critical thinking abilities, and ethical foundations to engage responsibly and constructively in social computing environments.

The evolution of social computing has not only revolutionized how teens today use social technology to communicate but also seek information, collaborate on projects, and explore their interests in digital spaces. As such, educators should consider adapting their pedagogical approaches to accommodate this relevant topic to connect students to the broader social implications of computing and guide them through the potential pitfalls of social computing such as misinformation, cyberbullying, and privacy breaches.

This research paper delves into the critical intersection of education and social computing for secondary-level students (ages 11--18), aiming to create sustainable teaching resources that can foster a community of socially responsible and aware individuals online. To assess the effectiveness of our resources, we used pre- and post-surveys following Bloom’s Taxonomy, a hierarchical model applied to assess comprehension levels of educational objectives through a variety of methods \cite{forehand2005bloom}. Through a comprehensive review of prior and related work, lesson development, and reflective school visits, this research paper seeks to spark conversation on the implementation of social computing education for K--12 students. In doing so, we aim to enable a generation of informed digital citizens to effectively shape the ever-evolving landscape of social computing.

%% file: my-files/related-work.tex
\subsection{Social Computing}
The rise in youth digital presence has produced polarizing effects for many secondary school students. Although youth value the accessible media digital spaces provide \cite{hassoun2023practicing, davis2013young}, studies show that they are often exposed to violence, harassment, misinformation, and other harmful content while online \cite{ito2013hanging, hassoun2023practicing, freed2023understanding}. As these risks coincide with computer science (CS) and behavioral science, social computing education proves to be especially relevant to youth, and in particular, secondary school students \cite{yardi2008hci, goldweber2013computer, jayathirtha2022supporting, fisher2018integrating, ko2020time}, making it a leading motivator for this work.

Prior research in the realm of social computing education has been predominantly centered around the college-level demographic, specifically exploring ways to enhance interest and retention in computing disciplines (e.g., \cite{isenegger2021understanding, buckley2008socially}). In contrast, our work seeks to elevate the discourse on social technologies, fostering critical thinking and understanding among a younger range of students. Our objective is to prompt students to question and understand the processes involved in the creation and utilization of technologies, as well as their broader societal impacts.

\subsection{Interdisciplinary Application}
Social computing overlaps with a variety of fields that require the practice of computational thinking. While CS education traditionally focuses on highly technical aspects, with limited work expanding into blended fields (e.g., \cite{amoussou2010interdisciplinary, carr2020interdisciplinary}), existing efforts emphasize social integration primarily within computing and programming courses. Other works often delve into integrating specific technical assignments and longer-term projects related to social topics into their respective courses. Rather than focusing on building technical skills through a social lens, we emphasize the high-level applications of social technologies through exploratory activities and discussion. This approach allows engagement from students interested in diverse interdisciplinary fields, encompassing humanities, science, mathematics, and career and technical education.

Our interdisciplinary approach goes beyond classrooms, reaching fairs, field days, and extracurricular activities. This unique method expands the impact of social computing education across diverse academic domains. To the extent of our knowledge, research has not been published with the integration of disciplines at a similar scope.

\subsection{Pedagogical Practices}
The importance of social computing education prompted us to practice culturally responsive pedagogy (CRP), a widely accepted paradigm in education \cite{ojeda2023describing, webb2018striving} embracing identities and experiences integral to students. CRP is shown to improve student engagement and learning outcomes \cite{ojeda2023describing, webb2018striving, stevens2022exploring, williams2017culturally}. Additionally, our curriculum follows an independent modular structure: each lesson addresses a different social computing topic with several learning objectives that are taught within one hour. While some studies follow a modulated format (e.g., \cite{wang2023virtual, meads2023moving}) and others use an independent lesson format (e.g., \cite{petelka2022principles, klassen2022run}), our curriculum takes advantage of the benefits behind both structures. Our design allows flexibility in lessons being taught consecutively for a comprehensive course or individually as they fit into existing curricula. During our lessons, we utilized four methods of evaluation: educator feedback, instructor observations, and student pre- and post-surveys, all of which are commonly implemented in CS education (e.g., \cite{carr2020interdisciplinary, meads2023moving, yardi2008hci, giacaman2023evolving, basu2020integration, petelka2022principles, cao2023experience, wang2023virtual}). This wide array of perspectives allowed us to effectively improve our lessons.

%% file: my-files/lessons.tex
This section provides context on the design process of our social computing modules and an overview of each module's content.

\subsection{Collaborators}
In order to create accurate, informative, and engaging content, we consulted with diverse perspectives in addition to our own.

{\itshape Students}. To guide the design of our materials, one of our coauthors is currently a high school student and another recently graduated from high school. These collaborators were able to provide insight from their personal experiences as individuals directly connected to the demographic this project seeks to impact.

{\itshape Researchers}. To assemble the specific social computing content and choose module topics, the remaining coauthors are social computing researchers. For specialized feedback on certain module topics, we consulted with researchers who work directly in our module topic fields. These collaborators provided accurate resources and expertise to help curate the social computing educational content.

{\itshape Professional Educators}. To ensure our materials are accessible and digestible to secondary school teachers, we received feedback on our lesson content and website design from educators who hosted us for school visits and a professional curriculum designer.

\subsection{Design}

Each of the six lessons follows an approximately one-hour format, designed for introductory levels with consistent content across all grades. With this structure, we aimed to prioritize factors like interactivity, reflection, and relevancy.

\begin{itemize}
    \item {\itshape Introduction}: Overview of the learning targets, a high-level background of the social computing topic, and examples.
    \item {\itshape Main Activity}: Interactive activity inquiring into one of the main themes and learning targets.
     \item {\itshape Reflection}: Class discussion on observations students made about the activity and its social implications.
    \item {\itshape Connection}: Class discussion connecting lesson activities and objectives to students’ everyday lives.
    \item {\itshape Closing Activity}: Interactive activity relating to another objective or theme, review of lesson content, and lesson wrap-up.
\end{itemize}

All activities were crafted to engage students in the ``Apply,'' ``Evaluate,'' and ``Create'' levels of Bloom's Taxonomy, fostering higher-order thinking skills.

\subsection{Content}
We created six independent lessons on various computing topics and focused our content on how they relate to social computing.

\subsubsection{{\itshape Data Management.}} With personal information easily accessible online \cite{moore2012digital}, we introduce the concept of a ``digital footprint'' with strategies to keep information safe on the internet. In the first activity, students work in small groups to decipher clues unique to their group. Their goal is to uncover information on a fictional character's life based on the pieces of information the character has publicly posted online. The class then assembles their clues to write a biography about the character. This activity demonstrates the vulnerabilities of publicizing information online. Then, students learn the ways digital footprints can be described and categorized. Finally, students apply this knowledge through the closing activity, where they match different data management scenarios to the various data categories.

\subsubsection{{\itshape Encrypted Messaging.}} As many youth use online messaging services to communicate with others \cite{schiano2002teen}, this module addresses the significance of encryption to keep information secure. The main activity is a class-wide simulation of asymmetric encryption. The class is divided into ``Senders,'' who send and receive written notes, and ``Deliverers,'' who deliver messages between Senders. The activity is broken into multiple rounds where a layer of security is added each time. After discussing the benefits of encryption from the activity, we introduce some concerns about encryption. In the closing activity, students use external resources and their peers to prepare for a class debate on the pros and cons of encryption.

\subsubsection{{\itshape Human-Computer Interaction (HCI) Careers}.} Because technology has become increasingly ubiquitous, this module focuses on careers related to HCI. In the main activity, students think of technology they want to redesign or create to address an issue of their choice. The design worksheet guides them through a process of brainstorming, connecting to the real world, physical design and functionality, and revision with peers. During the revision process, we emphasize the need for diversity in science, technology, engineering, and math (STEM) fields \cite{allen2014reimagining}. By encouraging students to share and iterate on their ideas with peers, we show the importance of perspective in the design process. Finally, the closing activity asks students to create career posters summarizing a STEM-related career of their choice and how it connects to social computing.

\subsubsection{{\itshape Machine Learning and Bias}.} Machine learning is rapidly being implemented in many everyday tools. However, machine learning tools contain a variety of underlying biases that can negatively impact users \cite{mehrabi2021survey}. In the main activity, students grade sample high school résumés and then compare their grades and criteria to an online artificial intelligence résumé screener. We reveal some real-life examples of how résumé screeners have shown identity bias in the past and reveal the potential job barriers this could cause \cite{dastin2022amazon}. In the closing activity, students explore text-to-image generation where they discover biases in some of the images generated firsthand. We provided guiding prompts that revealed bias in the generated images and encouraged students to try out their own text prompts as well.

\subsubsection{{\itshape Misinformation}.} There is an overwhelming amount of information consumed on a daily basis through online mediums, and it can be difficult to identify and filter misinformation \cite{wu2019misinformation}. The main activity is a class-wide simulation of how misinformation is spread. In this activity, we separate the students into “Users,” who consume media, and “Sources,” who are companies that produce and present media. Sources strategically choose from a wide pool of articles containing both factual and misinformation headlines to pitch to Users. Users then go around to hear pitches and pick their top three Sources to give their money to based on what interests them the most. After calculating which types of articles got the most money, it is typically found that the Sources that pitched misinformation gain the most money. After learning about why misinformation is spread and strategies on how to identify it, we test the students on how well they can identify misinformation. By using the skills that were taught, they go through a mix of headlines and discuss as a class whether they believe the article is misinformation or not.

\subsubsection{{\itshape Online Behavior}.} The secondary school age demographic is highly active on online platforms, thus this module explores the many beneficial and harmful effects online behavior has on youth \cite{lusk2010digital}. The main activity is a “Choose Your Own Adventure” where groups of students make different choices around the classroom that lead to different endings. As they make choices as a group, students discuss the potential ramifications of each option to guide them through scenarios resembling scamming, online harassment, academic integrity, and social comparison. In the closing activity, students inspect different packets of social media captions and pick the one packet they feel is the most appealing to them. Each packet contains one caption that focuses on a more negative experience in comparison to the others, which makes students reluctant to choose certain packets. This activity helps students understand that social media is often unrepresentative of the real world, especially as online users tend to primarily share the highlights of their lives.

%% file: my-files/visits.tex
We visited 13 different schools in our local area to pilot our lessons, reaching around 1,405 total middle and high school students. Since the end of 2022, we have conducted 24 visits across all schools.

\subsection{Outreach}
Two major announcements were sent to local secondary school teachers, expressing our willingness to teach social computing modules at schools. The first announcement in January '22 resulted in 10 visit requests, while the second in May '22 led to 38. The overwhelming response posed a challenge, as fulfilling all requests with limited people to teach was difficult. Ultimately, some visits had to be postponed to the next school year. Additionally, we collaborated with our university's K--12 Outreach Committee, conducting secondary school visits, participating in fairs and workshops during Computer Science Education Week, and presenting at regional educator networking events.

From our outreach efforts, we had 48 total school visit requests. Misinformation was our most requested topic with 33.30\% of teachers requesting it, followed by Machine Learning and Bias (27.10\%), Online Behavior (16.70\%), and Data Management (4.20\%). Encrypted Messaging initially received no teaching requests, but after consulting with teachers, we successfully piloted the module at two schools. In total, we delivered all six modules to both middle and high school levels at least once.

Among the 24 high school visit requests, Machine Learning and Bias (37.50\%) and Misinformation (29.17\%) were the most requested modules by teachers. For the 24 middle school requests, the most sought-after modules were Misinformation (37.50\%) and Online Behavior (37.50\%).

\subsection{Demographics}
All school visits were in Seattle Public Schools, the largest public school district in the state, where students are equipped with devices like laptops and tablets for classroom instruction. Based on survey results, 64.40\% of recorded students are in middle school (grades 6--8) and 35.60\% are in high school (grades 9--12). Out of those students, 50.30\% are Male, 43.70\% are Female, and the remaining 6.00\% self-identify as Non-Binary or another specified gender.

As it pertains to racial and ethnic diversity, we collected information from a wide range of backgrounds such as Black or African American (11.45\%), East Asian (13.99\%), Hispanic or Latino (10.24\%), Middle Eastern or North African (2.43\%), Native American or Alaska Native (1.98\%), Native Hawaiian (0.99\%), Pacific Islander (3.31\%), South Asian (5.07\%), Southeast Asian (11.34\%), and White (66.48\%). Students had the option to select more than one ethnic background.

\subsection{Pre- and Post-Surveys}

We administered optional pre- and post-surveys at the start and end of each lesson to assess students' social computing comprehension, module feedback, and demographics. Guided by Bloom's Taxonomy, our questions covered a spectrum of complexity, ensuring students grasped content through recognition, application, and creation.

Before the lesson started, we distributed the pre-survey. We asked students about their current knowledge of social computing and the specific social computing topic at hand through multiple-choice questions (MCQ) designed around each module's three learning targets. These questions focused on the ``Remember'' category of Bloom's Taxonomy, allowing students to demonstrate their ability to recall module concepts. We had 920 pre-survey responses across all school visits.

At the end of each lesson, we gave students time to complete the post-survey. We asked the same questions from the pre-survey in addition to a free-response question (FRQ) to demonstrate a deeper comprehension of the topic and a space to provide feedback on their experience with the lesson leveraging the ``Understand'' and ``Analyze'' levels of Bloom's Taxonomy. We had 629 post-survey responses across all school visits. We suspect that the difference in the amount of pre- to post-survey responses is due to limited time at the end of class which rushed students and a lack of energy from students to complete a longer survey after enduring a full lesson.

%% file: my-files/results.tex
\subsection{Comprehension Evaluation}
\begin{figure}[hb]
    \centering
    \includegraphics[width=\linewidth]{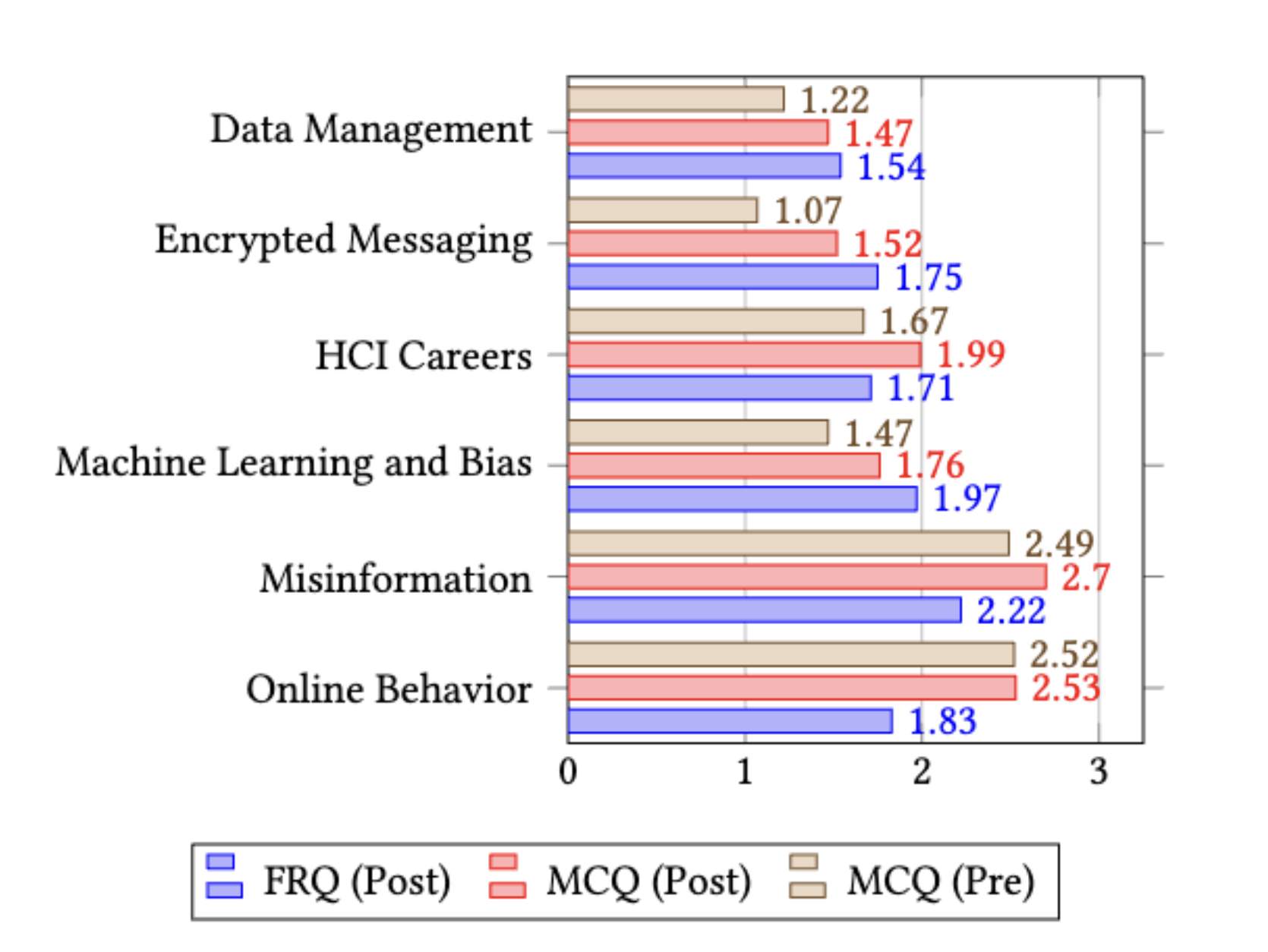}
    \caption{Student Pre- and Post-Survey Results by Lesson}
    \Description{Student Pre- and Post-Survey Results by Lesson}
    \label{fig:module_results}
\end{figure}

To quantify student comprehension, we gave students the same three MCQs on their respective module topics before and after the lesson. The post-survey included an FRQ broken into three parts. Both MCQs and FRQs were graded from a range of 0--3 based on correctness and demonstrated understanding, as seen in Figure ~\ref{fig:module_results}. Post-survey MCQ scores surpassed pre-survey results for all modules. For modules with closely aligned pre- and post-MCQ scores, it is likely attributed to higher percentages of students reporting having prior knowledge such as 70.42\% for Misinformation and 46.43\% for Online Behavior participants.

Through qualitative assessment, students made deep connections to the social computing topic they learned in their daily lives.
\begin{quote}
    Learning about misinformation today[,] I have realized that in my own life[,] if I am intrigued by something, especially something that is flashy and crazy like how misinformation tends to be[,] I will put my effort and money into said thing[,] even if it hurts me later on. (Misinformation Participant: Grade 12)
\end{quote}

\subsubsection{Understanding of Social Computing}

When students were asked to describe social computing and its impact on them, 50.60\% of students did not have a clear understanding in the pre-survey. In the post-survey, only 11.10\% of students were unsure and most were able to accurately connect social computing to the lesson topic. This student is one of many who showed a deep comprehension of the topic by describing its beneficial and harmful effects:
\begin{quote}
    I would describe [social computing] as online systems/platforms that allow us to communicate and have both positive and negative impacts. I think that social computing impacts me positively by letting me communicate with my friends and family, but I think it also impacts me negatively in that I sometimes compare myself to others and that probably lowers my self-esteem. (Online Behavior Participant: Grade 10)
\end{quote}

\subsection{Pedagogy Preferences}
To evaluate teaching practices that resonate most with students in accordance with learning social computing topics, we collected information on what helps them learn best. 70.20\% of students reported that demonstrations where instructors show concepts to the class and answer questions benefit them. Other significant practices noted were instructors working with smaller groups (39.80\%) and providing project-based assignments (38.50\%). We kept these preferences in mind as we iterated and improved our lesson content by incorporating opportunities for instructor demonstrations and small group projects.

\begin{figure*}[ht]
    \centering
    \includegraphics[width=\linewidth]{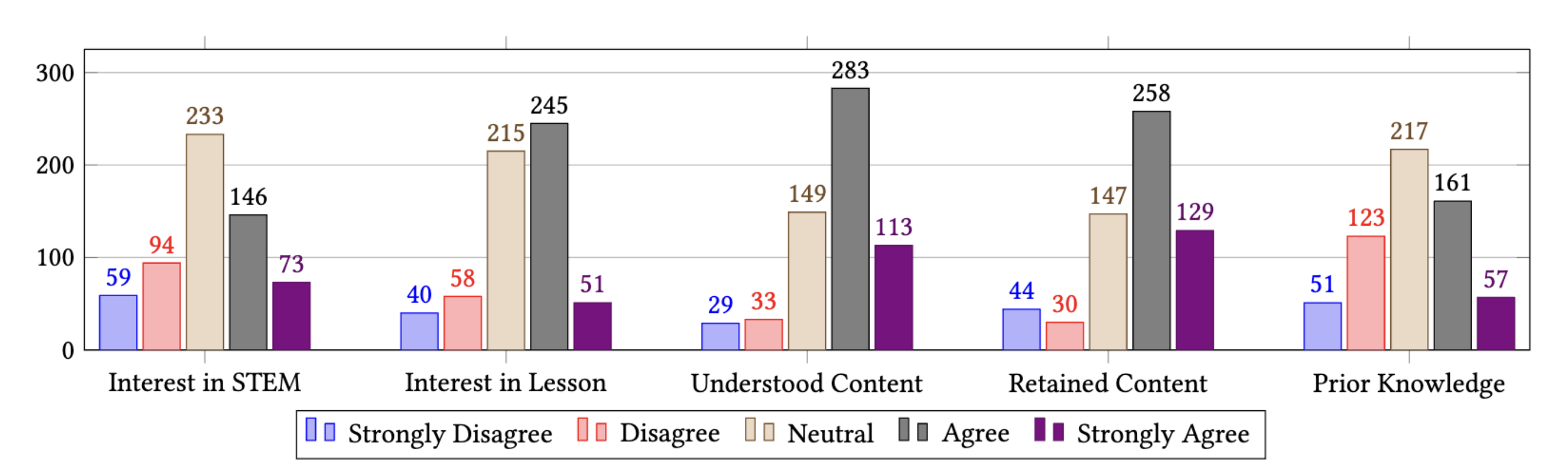}
    \caption{Student Self-Evaluation of Lesson Comprehension}
    \Description{Student Self-Evaluation of Lesson Comprehension}
    \label{fig:self_eval}
\end{figure*}

\subsection{Student Interest}
In addition to measuring comprehension, we collected information on students' self-reported interest and understanding of the lesson in Figure ~\ref{fig:self_eval} using a five-point scale of Strongly Disagree: 1 to Strongly Agree: 5. From this data, we discovered that 81.13\% of students expressed greater lesson interest compared to their STEM interest and 63.65\% learned new concepts after the lesson. The average change in the interest score from STEM to lesson was 0.22 with a standard deviation of 0.99. 

Furthermore, it was observed that the majority of students identifying with historically underrepresented minority groups in technical fields \cite{national2019women} demonstrated equal or increased interest in STEM post-participation. Specifically, 75.45\% of Female and Non-Binary students expressed positive sentiments, and among the Non-White and Non-Asian demographic, 75.68\% reported favorable views on STEM interest. These results highlight the effectiveness of our lessons in actively engaging and maintaining student interest in STEM.

%% file: my-files/teaching-resources.tex
After refining our materials through visits and incorporating feedback, our goal is to make them user-friendly for educators. Here, we outline the teaching resources we provide.

\subsection{Lesson Materials}
When developing each lesson, we created a lesson outline detailing content and timing, a slideshow with presenter notes, and related handouts. All of our lesson materials were created with Google Docs Editors, an effective office suite for educators that makes it convenient to download and remix for a teacher's personal goals and use \cite{romero2018improving}. Using this platform makes creating copies and personalizing our lessons fairly easy for educators.

\subsection{External Resources}
While developing our materials, we discovered various learning tools and resources relevant to our module topics. Our website acts as a centralized hub, listing related articles, online games, videos, and lesson plans. We aim to regularly update and expand these resources to make our website a comprehensive resource for social computing education.

\subsection{Website}
We created a user-friendly website, housing all resources for teachers. The website serves as a valuable tool for educators aiming to integrate our lessons into their curriculum. Find all materials posted at: \url{https://social.cs.washington.edu/sfl-curriculum/}.

\begin{figure}[h]
    \centering
    \includegraphics[width=\linewidth]{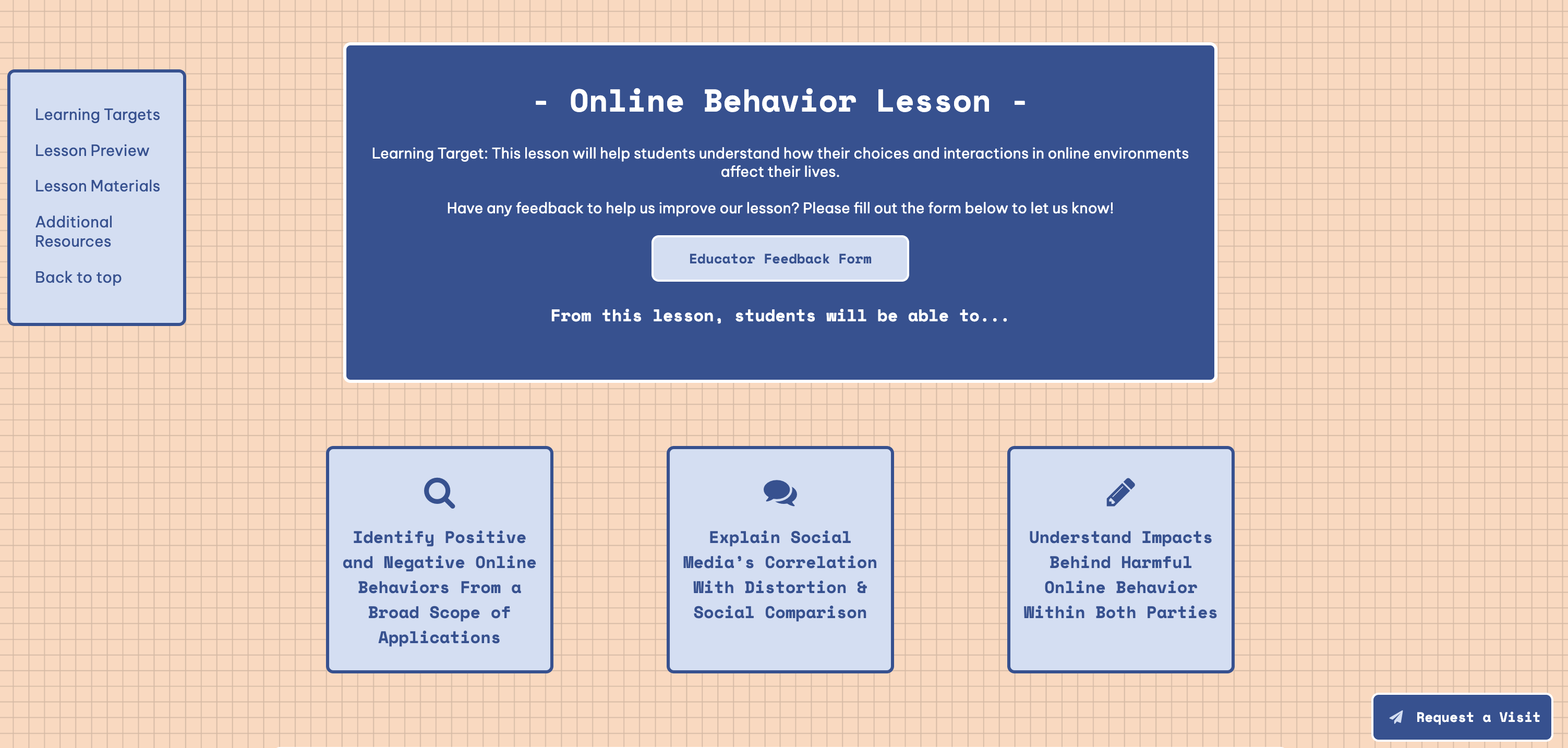}
    \caption{Section of the project website detailing lesson content, resources, and goals for the Online Behavior module}
    \Description{Section of project website detailing lesson content, resources, and goals for the Online Behavior module}
    \label{fig:website}
\end{figure}

In developing the site, we sought input from educators and researchers, refining it for optimal usability. Teachers reported increased confidence in presenting lessons after reviewing the materials. The website is structured with key sections: "Home" for an overview, "About Us" for project details, "Modules" for specific lessons (see Figure ~\ref{fig:website}), and "Contact" for visit requests and team details. We aim to continuously seek feedback from teachers and students to enhance our resources. Future plans involve expanding external resources on the website and creating additional support materials, such as example lectures and activity simulations, to aid teachers in independent lesson delivery.

%% file: my-files/reflection.tex
This section offers insights and recommendations from our experience in designing and teaching social computing lessons.

\textbf{Adaptability in lessons is key.} Since each lesson is separated into distinct sections with multiple activities, it is easy to adapt the lessons for a wide range of time lengths. Throughout all the school visits, we taught 30 to 120-minute class periods. For shorter class sessions, we cut down content by removing the closing activity or limiting class reflection and discussion. For longer classes, we gave ample time for class discussion or taught an additional module altogether. With the adaptability of our social computing modules, we were able to always make use of all the class time.

The lessons are adaptable to diverse class settings, accommodating 8-60 students based on factors like graduating students or shared lessons among teachers. The classrooms also varied in layout, ranging from large tables to rows of desktops. Despite varying layouts, our lessons consistently boosted comprehension, proving effective. For larger classes, we recommend multiple instructors or assistants to ensure focus, especially during year-end sessions when students may seem disengaged.

Our last form of adaptability is to accommodate independent or collaborative participation for students. All of our lesson's small group activities can be done individually if students prefer. During visits, a few students expressed comfort in doing certain activities alone instead of with peers. We observed that students were still able to actively learn and engage with the activity no matter what format they chose to participate in.

\textbf{Independent lessons allow for year-round flexibility.} When sending outreach emails to secondary school teachers district-wide, we had a significant amount of demand near the end of the year, in between semesters, or after Advanced Placement examinations. Many teachers emphasized the convenience of independent lessons by expressing how our lessons can easily be integrated during class on days they have more free time. All of the module topics are connected through the subject of social computing and can be taught consecutively to students. However, since the lessons do not build directly off one another, teachers have the freedom to choose individual lessons as they fit into their existing curriculum.

\textbf{Students connect social computing to their own identities and values when given the space to.} Many of our lesson activities were intentionally open-ended to allow for student self-expression. We found that students gravitated toward topics relating to their identities and interests when given creative prompts. This was particularly demonstrated in the main activity of our HCI Careers module, where most students designed technology aimed to help solve matters relating to their own cultural, ability, socioeconomic, and gender identities. From this observation, we highlight how social computing can directly tie into culturally responsive pedagogy.

\textbf{Social computing is applicable beyond computing courses.} To our surprise, most of our school visit requests were from teachers who did not teach computer science at their schools. Our visit requests were for a variety of class subjects such as English Language Arts (ELA), Psychology, Leadership, Health, Career Exploration, and all subjects of STEM.

Educators who hosted us detailed significant connections between our lesson modules and their current curricula, showing the importance of these topics within all academic subjects.
\begin{quote}
    Students in science get exposed to a lot of misinformation--this was especially exacerbated by the isolation and social media influx during the COVID-19 pandemic with regards to biology. (Misinformation Reviewer: High School Biology Teacher)
\end{quote}

The interactivity of our lessons allows for new depths of learning that complement the skills practiced in contrasting subjects.
\begin{quote}
    I do [no]t think it seems natural to be in an ELA class[,] but we did an argument unit about artificial intelligence [(AI)] and it helped to see the information presented in a different way with more examples so it was wonderful. We read about it[,] but it's a lot different to experience using an AI tool. (Machine Learning and Bias Reviewer: Middle School ELA Teacher)
\end{quote}

\textbf{Social computing is relevant to everybody.} Even if certain students are not necessarily interested in building technical skills or learning how specific technologies work, technology still affects their everyday lives. By introducing the concept of social computing, we focus on the broader social implications of technology that connect to students. This student expands on social computing and its omnipresence in their life:
\begin{quote}
    I did [no]t know the definition[,] but I have heard of the concepts. I would describe social computing as the idea and process of online information and data/communication digitally. It impacts me because I use the internet/digital platforms for entertainment, connecting with friends, and education, which means I am interacting with this digital web of computing daily. (Online Behavior Participant: Grade 10)
\end{quote}

Ultimately, there is an evident disparity in how social computing is introduced in secondary-level classes. Many students from our study may have been aware of certain computing topics but did not acquire formal insight or education from their existing curricula.

%% file: my-files/next-steps.tex
This project highlights the impactful influence of social computing education on youth. We aim to extend our reach by persisting in local school visits and expanding efforts to raise awareness of social computing topics beyond our initial target audience.

In hopes of reaching students beyond our local area, we plan to offer virtual alternatives to our activities. For further breadth of lesson options, we are exploring the addition of Recommender Systems and Mental Health modules. For further impact on teachers, we are looking into opportunities to host workshops for educators on how to teach this content and are promoting these materials further to organizations and educators. We encourage them to integrate these materials into their classes, focusing on our most sought-after modules, including Misinformation, Machine Learning and Bias, and Online Behavior. We see a great opportunity to expand these resources to a much younger and older demographic and encourage members of the computing education community to adopt social computing concepts in their work.